# COVID Down Under: where did Australia's pandemic apps go wrong?

Shaanan Cohney
*University of Melbourne*
shaanan@cohney.info

Marc Cheong
*University of Melbourne*
marc.cheong@unimelb.edu.au

*Abstract*—Governments and businesses worldwide deployed a variety of technological measures to help prevent and track the spread of COVID-19. In Australia, these applications contained usability, accessibility, and security flaws that hindered their effectiveness and adoption. Australia, like most countries, has transitioned to treating COVID as endemic. However it is yet to absorb lessons from the technological issues with its approach to the pandemic. In this short paper we a) provide a systematization of the most notable events; b) identify and review different failure modes of these applications; and c) develop recommendations for developing apps in the face of future crises. Our work focuses on a single country. However, Australia's issues are particularly instructive as they highlight surprisingly pitfalls that countries should address in the face of a future pandemic.

*Index Terms*—COVID-19, QR codes, contact tracing, Australia, security, uptake

## I. INTRODUCTION

In the wake of COVID-19, governments around the world turned to technology as a way to help prevent and track the spread of the virus. Naïvely one may have assumed that countries would coalesce around standardized and shared technology—much as they did around vaccines. However, the technologies adopted by different countries differed widely, with no clear consensus on which approaches were most effective. At least eleven independent frameworks were designed for contact tracing alone, and these varied in both the underlying technologies, the trust models, and in usability.

Australia's experience with COVID-tech is particularly illustrative and merits further exploration as a case study. Why? Among every country in the world, Australia imposed among the longest lockdowns in the world [1]—an event that the technology was supposed to help prevent. Studying the specific challenges and shortcomings that Australia faced with its COVID-tech can provide valuable insights and lessons should humanity (expectedly) be faced with further pandemics.

One of the primary technological measures implemented in Australia was the COVIDSafe app—a Bluetooth-beacon tracing app. However, the app not only failed to provide helpful data, it did so at a cost to government of more than ''$21 million'' [2]. Australian consumers raised concerns about the usability, security, and ethical implications of the COVIDSafe app and other technological measures implemented in the country. Frustration with Australia's applications boiled over, not only because of their lack of success in stemming the tide of the virus but because of independent issues caused by their introduction. In one high-profile instance Western Australia Law enforcement appropriated QR code check-in data [3] for non-virus related purposes—leading to public outcry.

Given the many challenges and failures of the Australia's COVID-related technology, it is important to take a step back and evaluate what can be learned from the experience. In this case study, we will examine the ecosystem of contact tracing and vaccine certificate applications within Australia's COVID-tech landscape. At a high level, we will explore how this ecosystem was fragmented by state-level decision-making and a rejection of international best practices in favor of a ''not-invented here'' mentality. It is important to note that we do not consider the broader ethical concerns surrounding contact tracing—this is well covered in other works by ethicists and NGOs including the ACLU's broadly cited whitepaper [4].

Our case study also *does not* detail how to develop cryptographically secure and context-appropriate contact tracing, nor does it provide a comprehensive design methodology for usability—as these are well-covered in other literature [5–7].

Rather, this paper serves as a retrospective. Through Australia's exemplar it highlights failure modes at the intersection of *security and usability* when governments rapidly-deploy novel technology in the face of disasters. While we present only one story, others have found echoes of Australia's issues in other jurisdictions [8–10].

The apps in our study fall broadly into three categories:
- *COVIDSafe*–the national contact tracing app, operating through Bluetooth Beacons on mobile devices
- QR-based check-ins for manual contact tracing, developed on a state-by-state basis. For completeness, we include a discussion on the fragmented third-party QR-based apps that were used before the emergence of *official* state-based apps.
- Vaccine certificate apps and (non-cryptographic) *digital signatures*

At time of writing, the situation in Australia has departed from the prior mandates and norms surrounding COVID-tech. Among these departures are the gradual phasing out of restrictions such as mandatory QR code *check-ins* and the sunsetting of contact tracing programs [2, 11, 12].

Accordingly, the contributions of this work are as follows:
- A systematization of the major Australian application types and their deployment/use history (Section II)

- Presentation of a number of security gaps in the application designs (Section III-A)
- A retrospective analysis of why the applications failed to achieve mass uptake or efficacy (Section III-B), along with our recommendations for future deployments (Section IV).

## II. Australian Use of Apps: 2020–2022

We begin by collating a history of Australia's use of COVID-related technology, necessary for our later analysis and compare it to successful efforts in Europe and the USA.

The COVID-19 response in Australia has seen marked shifts, from its initial outbreak in early 2020 to the implementation of long periods of lockdown in 2020 and 2021, to the current state of relaxed restrictions. As of today, Australian still faces sustained high case loads, but has few remaining measures to mitigate the spread among the general population.

To provide context for our case study, we have included a timeline of major events relevant to the use of COVID-related technology in Australia, in Figure 1. Of particular note are the developments in the states of New South Wales (NSW) and Victoria (VIC) which sustained the highest case loads [13], have the highest populations, and most severe restrictions over the duration of the pandemic. As a result, most of the lessons that readers can learn from the broader ecosystem can be extracted by focusing on these two states.

### A. Initial response: Bluetooth Beacon-based contact tracing

In the early 2020, researchers and governments scrambled to develop technological solutions for contact tracing: identifying the network of individuals who may have come into contact with a person who has tested positive for the virus. This technique has been used for decades to address infectious diseases, but the COVID-19 pandemic marked the first time that digital technologies were used on such a wide scale [23]. Technologists recognized that short-range wireless beacon traffic could serve as a strong proxy for determining which individuals had interacted, but the challenge then became how to turn this observation into a practical tool that could be widely adopted by the general population.

In Europe, consortia of researchers and governments developed a series of protocols including the two major competitors dp$^3$t [5], and PEPP-PT/PEPP. While the protocols and subsequent implementations received their fair share of legitimate criticism, the scope of collaboration and transparency helped to promote cost-effectiveness, continuous improvements, and public scrutiny of the applications.

Across the pond, Apple and Google each developed their own mobile *Exposure Notification (ENF)* APIs based on dp$^3$t. These APIs were designed to allow localities to quickly develop apps that could leverage privileged operating system (OS) functions to perform contact tracing. While these systems captured only the limited data Apple or Google permitted (less than what some governments desired), they benefited from access to low-level access to the OS. In the case of iOS, this allowed contact tracing to function even when the screen was off and the device locked— a feature not available to apps that were not using Apple's API [24].

In April 2020, the Australian federal government introduced the widely panned–and since abandoned–COVIDSafe contract tracing app [15]. The app was developed by the Digital Transformation Agency in conjunction with Boston Consulting Group at a cost of ''$10 million in developing the app ... $7 million on advertising and marketing, $2.1 million on upkeep and ... $2 million on staff.'' [25]

From its outset, civil society and the research community identified critical issues [26–28] with the app—while presenting viable alternatives. In particular COVIDSafe suffered from an iOS flaw, requiring some users to leave their phone unlocked and with its screen on, in order to use the app; this limits its utility in contrast to, say, the iOS ENF mentioned previously [24]. This limitation, which was denied by the government services minister [24, 29], reduced the app's utility compared to the iOS ENF.

In December 2020, the government switched to the open-source Herald Protocol [30] to ameliorate the many reported issues [31]. In justifying their decision to use Herald instead of the ENF, the government argued that the ENF focuses ''primarily on notifying individuals who may have been exposed to COVID-19'' while COVIDSafe ''was designed primarily to assist health officials to personally contact at-risk individuals'' [31]. The statement did not clarify the benefits of mediating contact through the government, rather than directly through the ENF.

In mid-2021, the Australian Department of Health published a report on the effectiveness of the COVIDSafe app and the Australian COVID tracking databases [32]. The report failed to provide substantive technical information on the app and took an uncritical approach to evaluating its efficacy. A second confidential and concurrent report was commissioned by the Department of Health [33], which has since been released, albeit being heavily redacted. Unlike the prior report, it was authored by independent consultants. Their findings were damning: ''the utilisation of COVIDSafe...resulted in high transaction costs for state contact tracing teams and produced few benefits'' [33].

Despite the initial hopes and efforts that went into its development, the COVIDSafe app ultimately proved to be a failure. By the time it was decommissioned, public reporting indicated that the app only discovered two positive cases and 17 close-contacts during the app's entire period of activity [25, 34].

While the government has not released details of the deliberations behind the application's development and deployment journalistic investigations, government reports, and independent analysis suggest that a number of key issues hindered its success. These issues include:

- Contact tracers required significant time to analyze the data outputs from the app
- The relatively low true-positive rate from the app compared to manual contact tracing
- Poor integration with system APIs caused issues with performance and efficacy
- Few users opted in to retention of their data for contact tracing (due to privacy concerns [33])

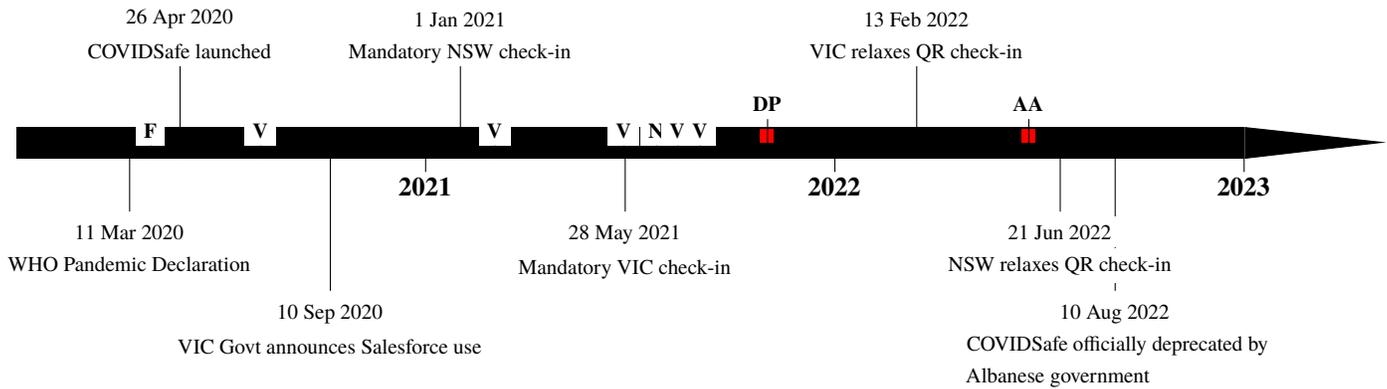

Fig. 1: Timeline of major events associated with technology-related measures to help curb COVID-19 in Australia. For context, dates of major leadership changes are included, reflecting policy changes (smaller squares: DP = Dominic Perottet in NSW; AA = Anthony Albanese at Federal level). Bigger squares on the timeline indicate start of major lockdowns (F=Federal, N=NSW, V=VIC), with the last major lockdowns ending in October 2021. [2, 14–22]

- Only a moderate proportion of the eligible population installed the app (≈30%), missing government targets

While the expenditure of approximately $21 million on the development and deployment of the COVIDSafe app may seem significant, it is important to contextualize it within the overall federal budget for COVID-remediation, which runs into the hundreds of billions of dollars. Despite this, our review suggests that the funds could have been deployed more effectively if the government had followed the best practices outlined in this paper. Further by collaborating with other countries and research groups, the government could have avoided many of the technical, usability, and ethical issues that ultimately led to the app's failure.

*B. Improvements: Decentralised, QR-based contact tracing*

While the federal government was busying itself with a high-tech approach to contact tracing, state governments and businesses independently adopted a mix of high-and-low tech approaches to contact tracing.

By around August 2020 [35], Australian businesses were independently collecting check-in data using a variety of ad-hoc methods. Contact tracers, employed by the states, collated data from locations suspected of having COVID cases, such as cafés and workplaces. These locations, in turn, collected data from customers using a wide range of technologies [36]: ranging from simple pen-and-paper notepads [35], to Google Forms[1], to food-ordering/restaurant booking apps [35].

However, by mid-2021 each state and territory transitioned to unified contact tracing systems [37] hosted on government service applications—ameliorating the challenges tracers faced with collecting data and interoperability.

State and territory governments in Australia each operate their own check-in applications and corresponding contact tracing infrastructure. However, not all systems are made the same. The states of Victoria and New South Wales, for example, use apps which also facilitate other e-government functionalities while the Australian Capital Territory, for example, uses a single purpose contact-tracing app (Table I).

In late 2021, the states of NSW (New South Wales) and Victoria added vaccine certificates to their government service apps. Visitors to businesses were required to scan a QR code for contact tracing, which would also display a graphic on their phone confirming their vaccination status. The announcements that accompanied the launch of these systems included claims that the systems were ''quite secure'' [46] because the applications included visual elements designed to prevent forgery.

Despite the fanfare, oversight in the design of the states'/territories' systems led to substantial issues. Criminals and anti-vaccine activists were able to easily exploit the weak security of the systems, issuing fake vaccine certificates that were difficult to distinguish from genuine ones [47].

Additionally, the use of check-in data for criminal investigations by law enforcement [3] damaged public trust in the systems, prompting privacy experts to caution against the secondary use of this data [48]. Furthermore, poor design choices in the Victorian QR codes made them difficult to use and led to low uptake [49].

The failure of the Australian government to engage with the critiques and concerns raised by researchers and the public regarding the implementation of vaccine certificates led to significant issues with the systems. A report by Nelson [50] highlighted the potential problems with the applications used between April 2020 and July 2021—yet the government failed to respond to any of these critiques.

Despite the challenges faced with the implementation of vaccine certificates in Australia, there were some positive developments as well. In particular, the federal government took a different approach with its vaccine passport for international travel, following criticism of the initial design [51]. While the initial design was subject to many of the same criticisms as the state systems, the version that launched used digital signatures to ensure authenticity—despite marketing materials that emphasized the ''Visual Digital Seal.'' The more effective scheme used QR codes to provide an easy interface that can accommodate both device-to-device communications and meatspace-to-device communications.

---

[1] From one of the authors' experience in Melbourne VIC, ca. 2020.

TABLE I: **State/Territory Check-In Apps.** Each state/territory developed their own check-in based contact tracing application. States/territories varied as to whether they integrated the functionality into an existing government services application or developed a new app whole-cloth. The minimum operating system requirements (Android and iOS) differ as well. Data sources: [38–45].

| State/Territory | App | Feature | Min. Android | Min. iOS |
|---|---|---|---|---|
| NT | The Territory Check In | Single-purpose | 5.0 | 12.0 |
| ACT | Check In CBR | Single-purpose | 5.0 | 12.0 |
| QLD | Check-in QLD | Single-purpose | 5.0 | 12.0 |
| TAS | Check In TAS | Single-purpose | 5.0 | 12.0 |
| WA | SafeWA | Single-purpose | 5.0 | 10.0 |
| SA | mySA GOV | Multi-use | 5.0 | 9.0 |
| NSW | Service NSW | Multi-use | 6.0 | 11.0 |
| VIC | Service VIC | Multi-use | 6.0 | 11.0 |

## III. Implications and Lessons

### A. Security

Australia's choice to advertise and design visual indicators of security—e.g., a 'green tick' for check ins—persistently came at the cost of strong cryptographic protections.

The fundamental design error is that the applications do not provide a reliable means for verifiers (typically businesses) to check the authenticity of certificates and check-ins. This stems from a lack of meaningful interactivity in the 'proving protocol': provers (guests to a business) briefly show an in-application graphic to the verifier, who has no meaningful opportunity to interact with the application or data to authenticate it.

As the proof is purely a graphic that is common to all authentic provers, an adversary merely must obtain a copy of the graphic and display it on their own device to fool the verifier. No matter the fancy pictures, bouncing syringe icons, or official government seals, UI elements can be replicated by any app with permission to display arbitrary content on a user's device—that is, most every app.

The alternative to this approach is interactive authentication: not only does the user present proof of vaccination (or check in), but a verifier checks the proof, establish its authenticity, and critically, whether it matches the user's identity. This is the technique used in the EU's vaccine passport system, and in the revision of the Australian federal vaccine passport system.

Although such systems inherently involve an additional step—wherein the verifier performs a meaningful check of the presented proof—a reasonable user experience can still be preserved, and the fundamental QR-design maintained. QR codes are more than capable of storing the requisite data for cryptographic verification. Paired with a modern signature scheme like BLS, the signature could occupy as little as 384 bits of a QR code [52]. This means smaller QR codes, and as a result, a smoother user experience.

### B. Hindrances to Uptake

It is possible to design a QR code system that is both secure and user-friendly. In fact, the schemes discussed earlier have been shown to be just as usable, if not more so, than some existing solutions that lack both security and usability. The usability desiderata for QR code systems include speed; compatibility; and accessibility. All three of these were at issue in a subset of Australian deployments [49].

*1) Speed:* The speed at which a consumer can scan a QR code depends on a number of factors, such as the physical size of the code, the error correction level, the amount of data stored, and the speed of any linked systems. Governments should have taken these factors into account when designing their QR code systems and made changes to improve usability. For example, the Victorian government's QR codes were excessively dense and slow to load for consumers due to their large size and bloated links. In comparison, QR codes used in Queensland and the ACT were smaller and faster to scan, despite using shorter links that stored the same amount of useful data. In our evaluation of the apps we found that earlier versions of Victoria's QR codes were even more complex, being ''almost 1000 per cent longer'' than those used in Queensland. By paying attention to these details, governments can create QR code systems that are both secure and user-friendly.

*2) Compatibility:* Beyond speed, the apps were targeted in particular at protecting the elderly—a demographic known to lag behind in using up-to-date devices [53].

As a result, backwards compatibility with the 'long-tail' of older devices is a more serious concern particularly with respect applications used for contact tracing. In Table I, we profiled the versions of the check-in software used on Android devices based on their Google Play Store listing, at time of writing[2]

A representative example is the December 2021 version of the Service VIC check in app which was limited to Android

---

[2]Where a current Google Play Store listing is not available, the Internet Archive Wayback Machine copy of the listing is consulted instead.

6.0 (released Q4 2015) and higher [54]. This would have been incompatible with between 1-2% of Android devices at the time [55]—likely disproportionately devices used by the elderly. This is a product both of app-store restrictions (whereby old API versions are deprecated), and choices made by app developers to use more modern features.

For the Apple iOS ecosystem (Table I[3]), multi-use apps fared better. Versions as early as iOS 9.0 (in the case of mySA GOV) are supported, compared to the iOS 12.0 required by their single-purpose app counterparts (NT/ACT/QLD/TAS). A requirement for iOS 12.0 and higher meant that any iPhone device from the iPhone 5S onwards (circa 2013) is supported; iOS 9.0 in SA's case equates to any device from iPhone 4S onwards (circa 2011). This provides support for more devices, up to 4 years older, than is the case with Android.

The lesson here, given our coverage of the 'minimum-supported platform version' for both major smartphone ecosystems, is that governments should ensure that as few users as possible are required to upgrade their phone to use the state-sanctioned check-in app. We revisit this point in our discussion (Section IV).

*3) Accessibility and Inclusion:* Accessibility issues extend beyond the elderly. In Australia, there is a significant digital divide that disproportionately affects certain groups. Research from the Australian National University found that "Aboriginal and Torres Strait Islander Australians, those born overseas in a non-English speaking country, those with low levels of education and those outside of the most advantaged areas'' [56] are less likely to use check-in apps.

This may be due to linguistic barriers that hinder usability, or to cultural responses to government mandates.

Furthermore, approximately 11% of Australians [57] do not have regular access to a smartphone, creating another group of individuals who may be excluded from using QR code systems. This exacerbates the impact of low uptake of QR code systems and highlights the need to consider accessibility in their design. It not only raises fairness concerns, but also poses safety risks, as pandemic regulations in certain states, such as Victoria and NSW, have barred individuals without check-in records or vaccine certificates from accessing public accommodations.

## IV. Discussion

Australia's roll-out of applications to support the pandemic response illustrates that if one compromises on security, privacy, or usability, this is reflected in the ultimate efficacy of the application.

Consumer apps operate in a environment with complex incentive structures. As some pundits have noted, getting the science or technology right is insufficient to solve complex policy problems, however, getting the technology wrong exacerbates deploying effective policy responses. While it is unclear that if COVIDSafe worked well, it would have delivered on the contract tracing needs, it is certainly the case that an application that goes unused provides a poor return on the substantial effort involved in its development, deployment.

This again highlights the importance of making usability a central pillar of developing apps for mass public adoption. However, it is also possible for the pendulum to swing too far, as with the ill-advised emphasis on 'visible digital seals' over cryptography—which still resulted in a broken socio-technical system. Australia's apps therefore teaching a lesson about the importance of combining usability with security and making *both* a design time priority. Here are several observations and recommendations based on extant literature as well as our observations on the Australian experience.

Below we recall several further issues which, while unsurprising at a high-level of generality, are reinforced by our findings and review.

**Closing the digital divide** Australia's apps disenfranchised many elderly and less technically sophisticated individuals—a result of both usability and technical decisions. Further deployments could consider better accounting for such populations by offering effective alternatives, such as wearable technologies [58, 59]. At a minimum, fallback options, such as manual forms and paper certificates, should be in place.

On the technical side, deployment of apps for critical services should not presuppose that a large proportion of the general public has the 'latest and greatest' smartphones out in the market. Having backwards compatibility with as many versions of mobile platforms and OSes as possible is a key desideratum for increasing the adoption of such apps amongst the general populace.

**Increasing usability** Australia's COVIDApps had severe usability issues (lags, failures, battery draining etc)—thwarting well-intentioned users at every turn. The users who successfully employed the apps were involved in *supererogation*—'going above and beyond' [60]—an usual behaviour on which it is unwise to rely.

This is particularly notable as contact tracing already relies in large part on social goodwill rather than enforcement. When no one is watching, will users still check in? If usability barriers make it more frustrating to check in, it is less likely that users will do so, thereby decreasing the system's efficacy [61].

**Common standards** Standardization has both cost (potential hindrances to innovation) and benefits (adoption of known good specifications). In Australia's case, a 'not invented here' mentality caused the government to forgo expertise in a technically complex domain—leading to a sub-par product. While it is unclear if adhering to a common standard, such as the Google/Apple APIs would have meaningfully improved pandemic outcomes, our work shows that the approach the government *did* take, certainly made that outcome less likely.

**Trust** Australia's reliance on social good-will makes trust a large factor in the extent to which users adopt contact tracing technologies [56, 59]. This is especially so where large amounts of private movement and activity data is concerned. In fact, trust is seen as a driver of adoption of the COVIDSafe app circa 2020; Biddle et al's [62] empirical findings suggest that ''...if the institutions involved in designing, delivering, and utilising data from the COVIDSafe app were more trusted, then a far higher percentage of people would have downloaded the app''.

---
[3]Where a current Apple App Store listing is not available, the Internet Archive Wayback Machine copy of the listing is consulted instead.

Our paper thus advocates for methods to gain public trust in Covid apps: ensuring transparency of the technology (such as open-sourcing the codebase for auditing and review); responding to public critics; providing clear standards for data governance; installing legal safeguards that the captured data will not be subject to out-of-scope use (such as for law enforcement as seen in the Introduction); and above all, investing in security infrastructure that follows better than industry-standard practices.

**Techno-Solutionism** The failure of Australia's COVID apps may be attributable to more than just bad design and management. The promises that accompanied the apps suggested that they would 'solve' the issues of viral spread. Much as Huesemann and Huesemann [63] suggest, misplaced face and hubris in the power of technology can redirect resources away from less technical approaches that may be more suited to a given challenge. This is not to say that technology is always wrong (as in the case of the stunning success of vaccines), rather a proposal to use technology in such critical situations should be carefully scrutinized to ensure it is the right tool for the job.

## References

[1] RMIT ABC Fact Check, ''Josh frydenberg says melbourne is the world's most locked down city. is that correct?'' *ABC News*, 2021 (cit. on p. 1).

[2] M. Butler, *Failed COVIDSafe app deleted*, en, https://www.health.gov.au/ministers/the-hon-mark-butler-mp/media/failed-covidsafe-app-deleted, Accessed: 2022-9-8, Aug. 2022 (cit. on pp. 1, 3).

[3] E. Laschon, ''Check-ins to SafeWA app unaffected after WA police accessed data as part of criminal investigations,'' *ABC News*, Jun. 2021 (cit. on pp. 1, 3).

[4] D. K. Gillmor, ''Principles for technology-assisted contact-tracing,'' *ACLU Whitepaper*, 2020 (cit. on p. 1).

[5] C. Troncoso *et al.*, ''Decentralized privacy-preserving proximity tracing,'' *arXiv preprint arXiv:2005.12273*, 2020 (cit. on pp. 1, 2).

[6] S. O. Blacklow, S. Lisker, M. Y. Ng, U. Sarkar, and C. Lyles, ''Usability, inclusivity, and content evaluation of COVID-19 contact tracing apps in the united states,'' en, *J. Am. Med. Inform. Assoc.*, vol. 28, no. 9, pp. 1982–1989, Aug. 2021 (cit. on p. 1).

[7] J. Bay *et al.*, ''BlueTrace: A privacy-preserving protocol for community-driven contact tracing across borders,'' *Government Technology Agency-Singapore, Tech. Rep*, vol. 18, 2020 (cit. on p. 1).

[8] L. Money, ''Strange code on your COVID vaccine record doesn't scan? California did that on purpose,'' *LA Times*, 2021 (cit. on p. 1).

[9] N. Singer, ''Virus-tracing apps are rife with problems. governments are rushing to fix them.,'' *NYTimes*, 2020 (cit. on p. 1).

[10] C. Kent, ''A comedy of errors: The uk's contact-tracing apps,'' *Medical Device Network*, 2020 (cit. on p. 1).

[11] T. Cowie and R. Eddie, ''Victorians abandon QR code check-ins as COVID cases drop,'' *The Age*, 2022 (cit. on p. 1).

[12] A. Smethurst and S. Ilanbey, ''Contact tracing scaled back as vaccination rate rises,'' *The Age*, 2021 (cit. on p. 1).

[13] Statista, *Number of covid-19 cases per 100,000 population in australia as of february 1, 2022, by state and territory*, https://www.statista.com/statistics/1103944/australia-coronavirus-cases-per-100-000-population-by-state/, 2022 (cit. on p. 2).

[14] World Health Organization, *WHO Director-General's opening remarks at the media briefing on COVID-19 - 11 march 2020*, en, https://www.who.int/director-general/speeches/detail/who-director-general-s-opening-remarks-at-the-media-briefing-on-covid-19—11-march-2020, Accessed: 2022-9-13, Mar. 2020 (cit. on p. 3).

[15] ABC News and K. Bourke, ''Australian government's coronavirus tracing app COVIDSafe downloaded 1 million times,'' *ABC News*, Apr. 2020 (cit. on pp. 2, 3).

[16] Parliament of Victoria Legislative Council Legal and Social Issues Committee, ''Inquiry into the victorian government's COVID–19 contact tracing system and testing regime,'' {Parliament of Victoria}, Tech. Rep. 193, Dec. 2020 (cit. on p. 3).

[17] NSW Government - Customer Service, *Covid safe check-in faqs*, https://www.clubsnsw.com.au/sites/default/files/2020-12/COVID%20Safe%20Check%20In%20-%20FAQs%20for%20businesses_201224.pdf, 2021 (cit. on p. 3).

[18] Victorian Government - Premier of Victoria, *New fine for businesses flouting qr code requirements*, https://www.premier.vic.gov.au/new-fine-businesses-flouting-qr-code-requirements, 2021 (cit. on p. 3).

[19] ''Victorian government eases some QR check-in and density measures but masks to remain for now,'' *ABC News*, 2022 (cit. on p. 3).

[20] NSW Government Digital Channels, *Check-in and QR codes*, https://www.nsw.gov.au/covid-19/business/check-in, 2022 (cit. on p. 3).

[21] J. Boaz, ''Melbourne passes buenos aires' world record for time spent in covid-19 lockdown,'' *ABC News*, 2021 (cit. on p. 3).

[22] A. Xiao, H. Tatham, R. Hayman, and C. Hanrahan, ''Greater sydney marks 100 days of covid-19 lockdown,'' *ABC News*, 2021 (cit. on p. 3).

[23] A. M. Brandt, ''The history of contact tracing and the future of public health,'' *American Journal of Public Health*, vol. 112, no. 8, pp. 1097–1099, 2022, PMID: 35830671 (cit. on p. 2).

[24] J. Taylor, ''Covidsafe app is not working properly on iphones, authorities admit,'' *The Guardian, Australia*, 2020 (cit. on p. 2).

[25] M. Butler, *Failed COVIDSafe app deleted*, https://www.health.gov.au/ministers/the-hon-mark-butler-mp/media/failed-covidsafe-app-deleted, 2022 (cit. on p. 2).

[26] C. Culnane, E. McMurtry, R. Merkel, and V. Teague., *Tracing the challenges of COVIDSafe*, https://github.com/vteague/contactTracing, Accessed: 2020-5-7, Apr. 2020 (cit. on p. 2).

[27] J. Taylor, ''How did the Covidsafe app go from being vital to almost irrelevant?'' *The Guardian*, May 2020 (cit. on p. 2).

[28] K. Leins, C. Culnane, and B. I. P. Rubinstein, ''Tracking, tracing, trust: Contemplating mitigating the impact of COVID-19 through technological interventions,'' *Med. J. Aust.*, p. 1, 2020 (cit. on p. 2).

[29] Prime Minister, Minister for Health, Minister for Government Services, and Chief Medical Officer, *Covidsafe: New app to slow the spread of coronavirus*, https://web.archive.org/web/20220129222705/https://www.pm.gov.au/media/covidsafe-new-app-slow-spread-coronavirus, 2020 (cit. on p. 2).

[30] Herald Project Contributors, *Herald project*, https://heraldprox.io, 2022 (cit. on p. 2).

[31] Digital Transformation Agency, *Covidsafe uses the Herald Protocol to improve app performance*, https://web.archive.org/web/20220303171017/https://www.dta.gov.au/news/covidsafe-uses-herald-protocol-improve-app-performance, 2020 (cit. on p. 2).

[32] Commonwealth of Australia as represented by the Department of Health, ''Report on the operation and effectiveness of COVID-Safe and the National COVIDSafe Data Store,'' Tech. Rep., 2021 (cit. on p. 2).

[33] Abt Associates and BNDA, ''Final Report - Evaluation of the Operation and Effectiveness of COVIDSafe and the National COVIDSafe Datastore,'' Tech. Rep., 2021 (cit. on p. 2).

[34] P. Karp, *Australia retires $21m CovidSafe contact-tracing app that found just two unique cases*, 2022 (cit. on p. 2).

[35] J. Taylor, ''Privacy concerns over Australian businesses collecting data for covid contact tracing,'' *Guardian*, Aug. 2020 (cit. on p. 3).

[36] Y. Murray-Atfield, ''Victoria moves to single QR code COVID-19 check-in system amid concerns about compliance,'' *ABC News*, May 2021 (cit. on p. 3).

[37] K. Curtis and D. Crowe, '''clear failure': Victoria lags other states on single QR check-in system,'' *The Sydney Morning Herald*, May 2021 (cit. on p. 3).

[38] Google Play Store, *The Territory Check In*, https://play.google.com/store/apps/details?id=au.gov.nt.health.checkin, 2022 (cit. on p. 4).

[39] Google Play Store, *Check In CBR*, https://play.google.com/store/apps/details?id=au.gov.act.health.checkin, 2022 (cit. on p. 4).

[40] Google Play Store, *Check In QLD*, https://play.google.com/store/apps/details?id=au.gov.qld.checkin, 2022 (cit. on p. 4).

[41] Google Play Store, *Check in TAS*, Archived copy at https://web.archive.org/web/20220106140213/https://play.google.com/store/apps/details?id=au.gov.tas.checkin, 2021 (cit. on p. 4).

[42] Google Play Store, *SafeWA*, https://play.google.com/store/apps/details?id=au.gov.wa.health.SafeWA, 2021 (cit. on p. 4).

[43] Google Play Store, *mySAGOV*, https://play.google.com/store/apps/details?id=au.gov.sa.my&hl=en&gl=US, 2022 (cit. on p. 4).

[44] Google Play Store, *Service NSW*, https://play.google.com/store/apps/details?id=au.gov.nsw.service, 2022 (cit. on p. 4).

[45] Google Play Store, *Service Victoria*, https://play.google.com/store/apps/details?id=au.gov.vic.service.digitalwallet.citizen, 2022 (cit. on p. 4).

[46] ''Vaccine certificates explained: how to prove your vaccination status as Victoria prepares to reopen,'' *ABC News*, 2021 (cit. on p. 3).

[47] S. F. Koob, '''they will become more desperate': Vaccination fraudsters shut down,'' *The Age*, Oct. 2021 (cit. on p. 3).

[48] A. Taylor, '''this must not be permanent': Privacy experts sound alarm over QR codes,'' *The Sydney Morning Herald*, Apr. 2021 (cit. on p. 3).

[49] L. Mannix, ''Victoria's QR codes badly made, developers say,'' *The Age*, Jun. 2021 (cit. on pp. 3, 4).

[50] R. Nelson, ''iOS Application background behaviour,'' 2020 (cit. on p. 3).

[51] J. Purtill, ''COVID-19 vaccination certificates at risk of forgery after discovery of another security flaw,'' *ABC News*, 2021, Accessed: 2021-12-9 (cit. on p. 4).

[52] D. Boneh, S. Gorbunov, H. Wee, and Z. Zhang, ''Bls signature scheme,'' Technical Report draft-boneh-bls-signature-00, Internet Engineering Task Force, Tech. Rep., 2019 (cit. on p. 4).

[53] B. B. Neves and G. Mead, ''Digital technology and older people: Towards a sociological approach to technology adoption in later life,'' *Sociology*, vol. 55, no. 5, pp. 888–905, 2021 (cit. on p. 4).

[54] Google Play Store, *Service Victoria*, https://web.archive.org/web/20210307034358/https://play.google.com/store/apps/details?id=au.gov.vic.service.digitalwallet.citizen, 2021 (cit. on p. 5).

[55] *Mobile & tablet Android version market share Australia*, https://gs.statcounter.com/android-version-market-share/mobile-tablet/australia/#monthly-202005-202105, 2021 (cit. on p. 5).

[56] N. Biddle, M. Gray, and K. Sollis, ''The use of QR codes to identify COVID-19 contacts and the role of data trust and data privacy,'' The ANU Centre for Social Research and Methods, Tech. Rep., Nov. 2021 (cit. on p. 5).

[57] Deloitte, ''Mobile nation 2019: The 5G future,'' Deloitte Access Economics, Tech. Rep., 2019 (cit. on p. 5).

[58] G. Grekousis and Y. Liu, ''Digital contact tracing, community uptake, and proximity awareness technology to fight COVID-19: A systematic review,'' en, *Sustain Cities Soc*, vol. 71, p. 102 995, Aug. 2021 (cit. on p. 5).

[59] A. T.-Y. Chen and K. W. Thio, ''Exploring the drivers and barriers to uptake for digital contact tracing,'' en, *Soc Sci Humanit Open*, vol. 4, no. 1, p. 100 212, Oct. 2021 (cit. on p. 5).

[60] D. Heyd, ''Supererogation,'' in *The Stanford Encyclopedia of Philosophy*, E. N. Zalta, Ed., Winter 2019, Metaphysics Research Lab, Stanford University, 2019 (cit. on p. 5).

[61] J. Joo and M. M. Shin, ''Resolving the tension between full utilization of contact tracing app services and user stress as an effort to control the COVID-19 pandemic,'' en, *Serv. Bus.*, vol. 14, no. 4, pp. 461–478, Dec. 2020 (cit. on p. 5).

[62] N. Biddle, B. Edwards, M. Gray, M. Hiscox, S. McEachern, and K. Sollis, ''Data trust and data privacy in the covid-19 period,'' The ANU Centre for Social Research and Methods, Tech. Rep., 2020 (cit. on p. 6).

[63] M. Huesemann and J. Huesemann, *Techno-fix: why technology won't save us or the environment*. New Society Publishers, 2011 (cit. on p. 6).